# Be-Educated: Multimedia Learning through 3D Animation

Zeeshan Bhatti[1*], Ahsan Abro[1], Abdul Rehman Gillal[2] and Mostafa Karbasi[3]

[1]Institute of Information and Communication Technology, University of Sindh Jamshoro,
[2]Universiti Teknologi Petronas, Malaysia
[3]Department of Information Systems, International Islamic University Malaysia

*Zeeshan.bhatti@usindh.edu.pk, abro400@gmail.com, a.rehman_g02792@utp.edu.my, mostafa.karbasi@live.iium.edu.my

*Abstract:*

Multimedia learning tools and techniques are placing its importance with large scale in education sector. With the help of multimedia learning, various complex phenomenon and theories can be explained and taught easily and conveniently. This project aims to teach and spread the importance of education and respecting the tools of education: pen, paper, pencil, rubber. To achieve this cognitive learning, a 3D animated movie has been developed using principles of multimedia learning with 3D cartoon characters resembling the actual educational objects, where the buildings have also been modelled to resemble real books and diaries. For modelling and animation of these characters, polygon mesh tools are used in 3D Studio Max. Additionally, the final composition of video and audio is performed in adobe premiere. This 3D animated video aims to highlight a message of importance for education and stationary. The Moral of movie is that do not waste your stationary material, use your Pen and Paper for the purpose they are made for. To be a good citizen you have to Be-Educated yourself and for that you need to give value to Pen. The final rendered and composited 3D animated video reflects this moral and portrays the intended message with very vibrant visuals.

## I. INTRODUCTION

In the current technological era, the schooling and learning process have gone beyond the conventional teaching methods. Use of multimedia contents is measured as a highly favourable and relatively obligatory. This gives a birth to principle of multimedia learning. Multimedia learning deals with using multiple media contents, such as Text, Images, Video, Audio and Animation, for teaching and learning purpose. Multimedia technology is now being used at high scale throughout the world. For example, most of people do not read books they use internet for knowledge. People now a days do not tend to read newspapers, instead they watch news on T.V channel. Hence, we decided to spread the message through 3D Animation and using the principle of multimedia Learning. Through this project, an approach to combine the cognitive multimedia learning principle with Animation based Learning techniques is designed. Generally, multimedia learning involves, images and video along with text, still 3D animation techniques are employed to attain a level of sophistication through which a strong message and lesson on education can be portrayed. Various studies show that every student can gain knowledge and understanding from a multimedia based simulated environment [1]. They can also apply that knowledge and information in a real world scenarios [2]. Therefore, this study aims to use multimedia learning approach to develop an animated video showcasing a message that would be easily understood and grasped by the ordinary students. The name of this 3D animated movie is Be-Educated. Which is basically a message to the youth and the students of Pakistan who don't get involved in studies and most of these students don't care about their stationary items such as Pen, paper and other things. The animation is developed using 3D graphics, a famous graphic's technique. Moreover, 3D graphics is being used to make games, cartoons and animated movies. The remaining of the paper consists, highlighting the previously done similar work in Section 2. In Section 3, various tools and technologies used in this paper has been addressed. The detailed developmental methodology and techniques are discussed in Section 4, whereas the final results are presented in Section 5. The paper is concluded with future directions in Section 5, and in Section 6 the external contributors to this project have been

*Corresponding Author: Zeeshan.bhatti@usindh.edu.pk

13

acknowledged, with list of reference articles given in section 8.

## II. SIMILAR WORK

Since the advent of Internet, the teaching and learning tools and methodologies have become immensely modernized not only at homes but also at schools. The use of multimedia contents has become a norm of fundamental learning mechanism for students. It has been discussed and proven by many researchers that animation can encourage and enhance the ability of the learner and viewer to understand and gain the message. Especially, when it has been used within the principles of cognitive theory of multimedia learning [3][4][5][6]. The basic definition of cognitive theory of multimedia learning is grounded on three core conventions. First assumption, for processing of information two separate - Audio and Visual data, channels are to be used. Whereas, second assumption says that the channel capacity will always be limited, exposing the fact that there will be only a partial and restricted amount of information that can be actively processed at any single given time. The last and third convention states that the **learning** is an active process concerning and dealing with filtering, selecting, organizing, and integrating information [3]. Figure 1 illustrates the basics of cognitive theory of multimedia learning.

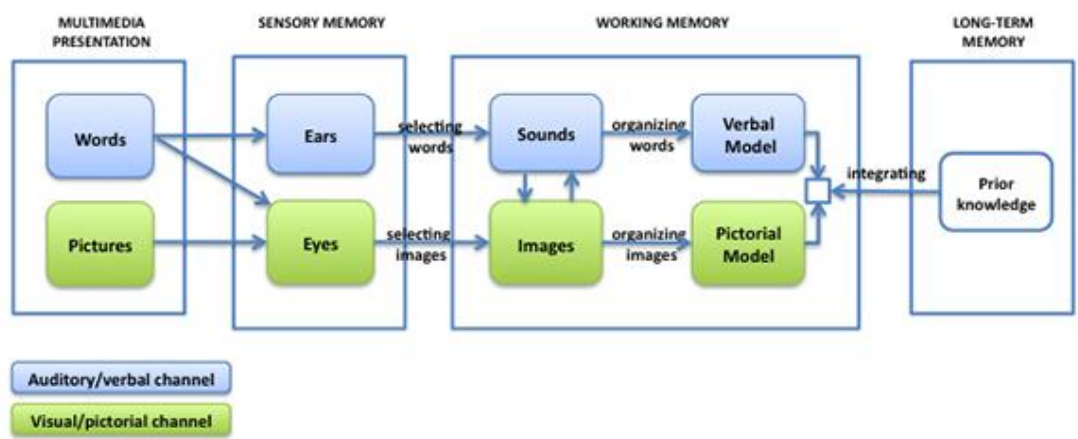

**Figure 1: Summary of Cognitive theory of Multimedia Learning (Source: [3])**

Various applications and courses have been specifically designed to cater for interactive multimedia learning approach. According to Rosenberg, for the purpose of teaching students about pedestrian safety, the best and cost-effective method is to use interactive multimedia learning approach [1]. Similarly, McCommas et al., uses Virtual Reality (VR) to help and teach students of primary schools to learn about the road safely and various ways in which to cross the street and crossing crossroads [2]. Whereas, Ann et al., uses interactive animation and videos to educate about traffic safety, highlighting the importance of interactive multimedia as a promising approach for such kind of teachings [7]. Similarly, Ravi et al., discusses a novel approach of developing an interactive multimedia application called "FIQIR Road Safety" for educating children of Malaysia about road Safety, using a framework comprising of multimedia elements, modules, learning theories and tests for usability acceptance [8] . However, Ahmed and Janghel design and develop a 3D animation to highlight and educate the people about the importance of not to Drink and Drive, using 3D Studio max to model and animate their content [9].

## III. TOOLS AND TECHNOLOGY

Several tools are used to develop this project, based on multimedia learning using the cognitive theory and principles of Audio and Visual Effects with limited channel capacity. The following are the tools and use of them.

- Adobe Photoshop (for making textures and using color scheme)
- Adobe Premier (for editing the video and some effects as final touches)
- Adobe Audition (for audio mixing, cutting and adjusting according to the scenes)
- Autodesk 3D Studio Max (Modelling all the characters, scenes, interior, exterior, Lightning, Cameras, Texturing on Models and Animation)

### 3.1 ADOBE PREMIER

Adobe Premier is used for the editing of video and merging scene together to make one complete scene or cinematic. Adobe Premier is based on Layer system working interface. Adobe Premier also has option for audio effects and separate layers for audio editing [10]. We have used Adobe Premier for the merging scenes and some video effects like " dip to black " and " dip to white " .

Final rendering and post processing is done in Adobe Premiere as all the scenes were put together in Adobe Premier, which is an video editing software. There were 130 layers and 13100 frames, which produced 7 minutes and 56 seconds animation. Started titles and ending credits also were made in Adobe Premier.

### 3.2 ADOBE AUDITION

Adobe Audition is basically used for the broadcast engineering, because it has some special effects and easy to manage the wave length of audio which helps in broadcasting at 192kbps [11].

For sound editing, Abode Audition software is used, for cutting the voices of characters, and made them into cartoonish voice to give a funny cartoony feel in the movie. Some background music effects used in this project, have been extracted from original movies sound tracks. Adobe Audition also has a Layer based working interface, which provides a user friendly environment. Every voice layer has a separate working layer within the timeline, and it is very simple to cut and make your own part from the music. For instance, just select the part of wave length, copy, make new file, paste, and save as in your required format.

### 3.3 3D STUDIO MAX

Autodesk 3Ds Max is a very powerful and advance software for making 3D objects. Its primarily used for modelling 3D architectural buildings, interiors designing, low poly game characters, locations and environment modelling [12][13]. However, the use of 3Ds Max for animated movie is still very limited. There are numerous tools and modifiers available in this software to easily and efficiently model any object imaginable. Therefore, for this project everything is constructed in 3Ds Max: Character designing, Interior Designing, Exterior Designing, Lightning, Animation, Camera angles.

## IV. DEVELOPMENT METHODOLOGY

The project development pipeline involves several stages and key components from script to Story boarding, Art design to 3D Modelling and Animation. This entire pipeline has been illustrated in figure 2, where each key phase of the project and flow is given.

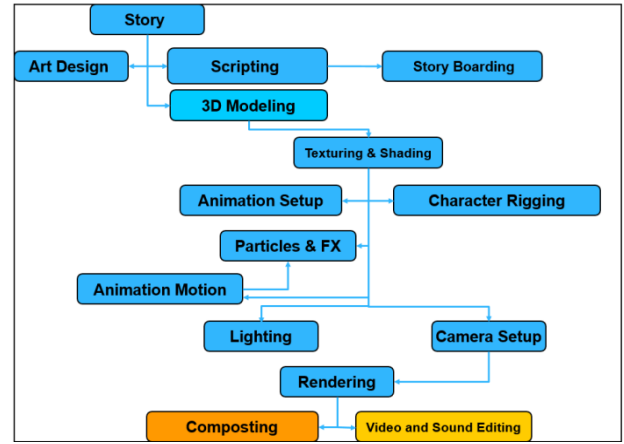

**Figure 2: Pipeline for 3D animated film production**

### 4.1 MODELLING A POLICE MARKER CHARACTER

The project development started with modelling of all the characters, locations, objects, props to be used in the story. Each character has been designed specially to look like a stationary object such as pen, pencil, eraser, and sharpener. Special techniques of character modelling based on extrude and pull, had been used to model them. The character of a Police man was designed as a Marker Pen.

To model the Police Marker character, a Polygon cylinder object was used in 3D Studio Max as shown in figure 3, where a Cylinder is taken from the command panel, and then it is converted into editable poly. The top faces of editable poly object is then selected and extruded from the top. In figure 4, a cap has been added by taking another cylinder using same techniques as discussed, converting the cylinder into editable polygon and extruding its faces to create a cap shape.

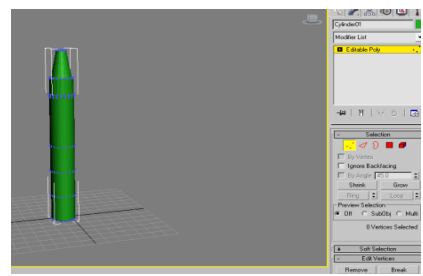

**Figure 3: Editable poly cylinder object**

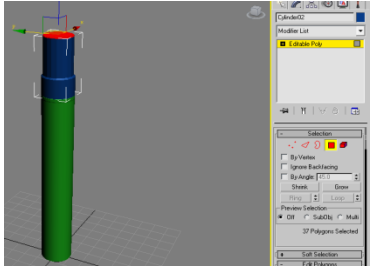

**Figure 4: Cap of Marker using poly cylinder**.

In the next step, the arms of the marker are modelled from cylinder and then convert into mesh smooth as shown in figure 5. For legs, the same procedure as followed for arms, take a cylinder, modelled the shape of legs, and then convert into mesh smooth as shown in figure 6.

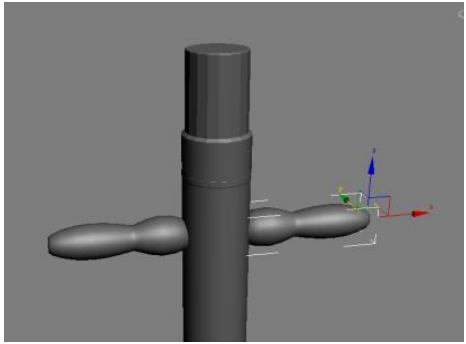

**Figure 5: Arms of Marker character**

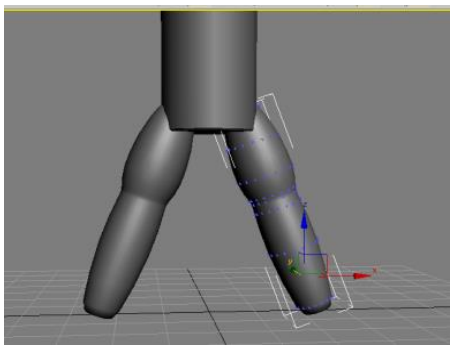

**Figure 6: Legs of Marker Character**

In the following step the Cap of Police officer is modelled, again by taking polygon cylinder object, then applying mesh smooth modifier on it. The front vertices of the mesh object are extruded outward to give the shape of cap as shown in figure 7. Finally, the eyes and lips are modelled next as shown in Figure 8 where the eyes are modelled using polygon sphere objects, whereas the lips have been modelled by using the line tool and extruding the line shape.

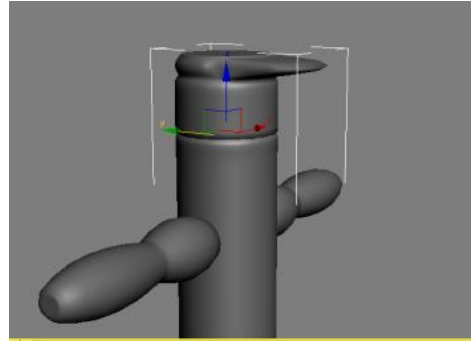

**Figure 7: The Cap of Police Marker Character**

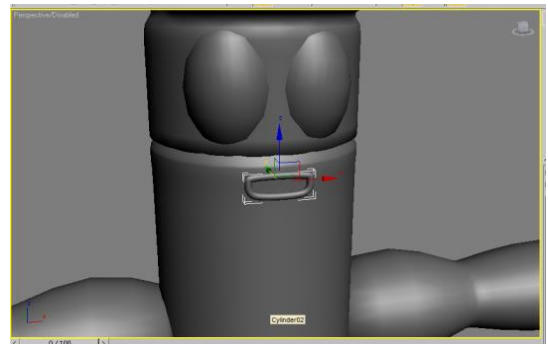

**Figure 8: Eyes and Lips of Marker Character**

The hands are modelled by taking polygon sphere, then extruding the fingers by selecting particular polygon faces. The hand polygonal object is then converted into a smoothed object through mesh smooth modifier. Finally, both the hands are attached to the arms as shown in figure 9. The shoes object has been modelled using a polygon box and mesh smooth conversion after modelling it into a show shape as shown in figure 10.

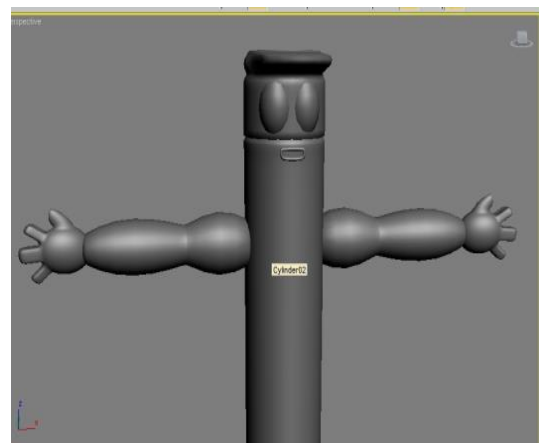

**Figure 9: hands attached to arms.**

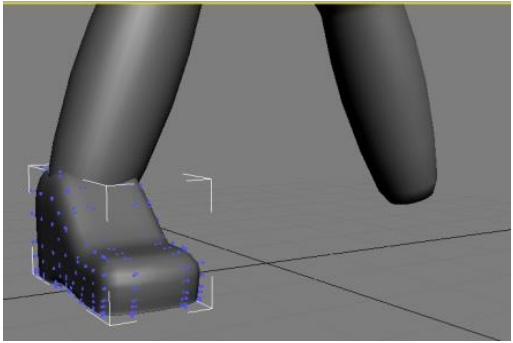

**Figure 10: Shoe modelled from poly box**

Figure 11 shows the final completed character of Police Marker. Similarly, all other characters have been modeled: the judge character as shown in figure 12 and eraser character in figure 13.

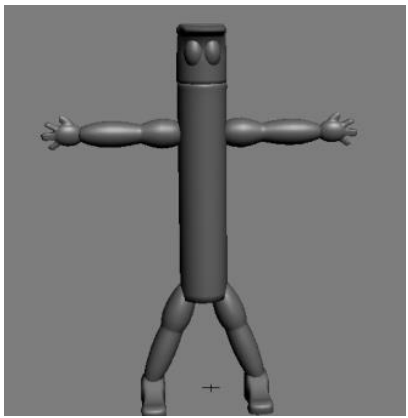

**Figure 11: Final Police Marker Character**

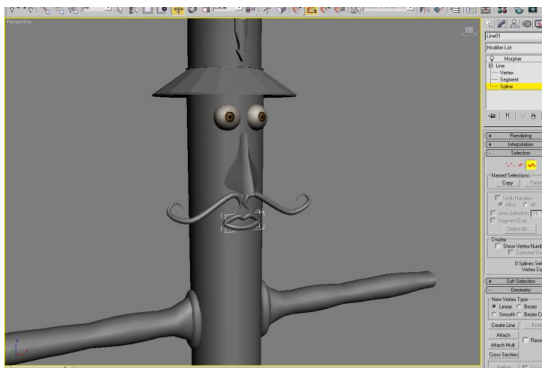

**Figure 12: The Judge Pen Character**

*4.2 RIGGING OF 3D CHARACTERS*

Rigging is the most critical part of character animation involving several complex steps and procedures. Animation is a process of bringing virtual objects to life, to do this, an animator requires a collection of controls and manipulators, which would enabled them to control the each part of character and animation [14].

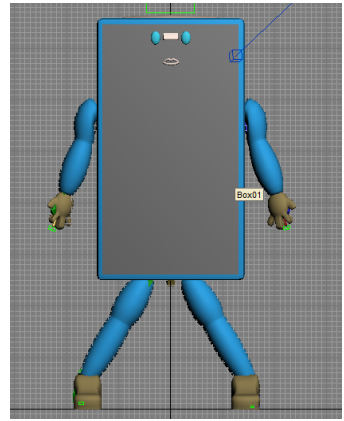

**Figure 13: The Eraser character**

The tediousness of physically doing this procedure for every character for every single frame, makes this process of animating a 3D character a time-consuming, difficult and problematic [15] [16]. Therefore to simplify this tedious process of rigging of various stationary characters used in this project, a built-in Biped character setup is used as shown in 14. In this figure 14, it is shown that this character will work as a human, by adding bones in the mesh body, applying biped which is the default human behaviour system in 3D max as shown in figure 15

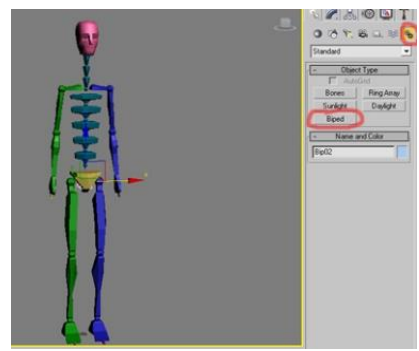

**Figure 14: Biped character Setup**

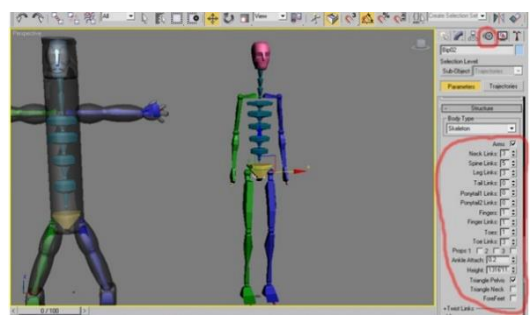

**Figure 15: Biped applied on a marker character**

To attach this biped to the modelled character, a built-in modifier, Physique is used as shown in figure 16.

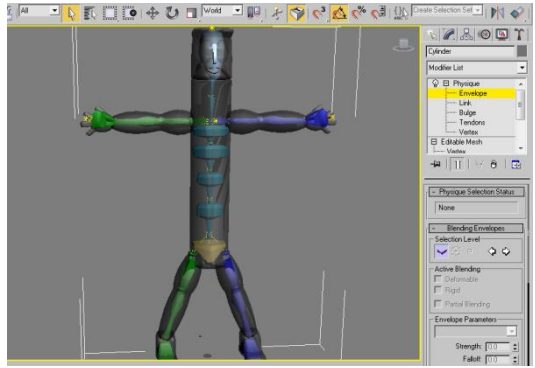

**Figure 16: Physique Modifier applied to attach model with biped.**

The components of the physique modifier were then modified to ensure correct attachment of skin. The envelope component around each bone part of physique modifier is scaled and adjusted to best suit the body size. Figure 17 shows hand joints and Figure 18 for foot joints, where the falloff of each envelope is adjusted around each bone.

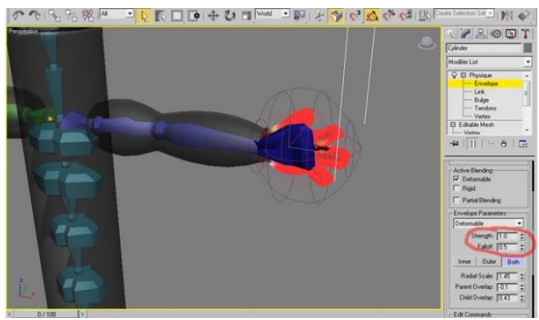

**Figure 17: Adjusting envelope of wrist joints**

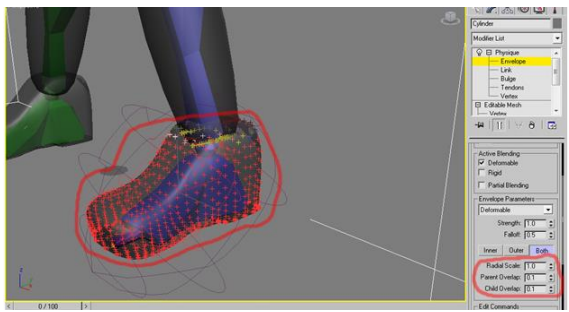

**Figure 18: Adjusting envelopes of foot joints.**

Similarly, all other characters were fitted with biped character setup for the rigging and physique modifier was used to attach the skin with bones as shown in figure 19.

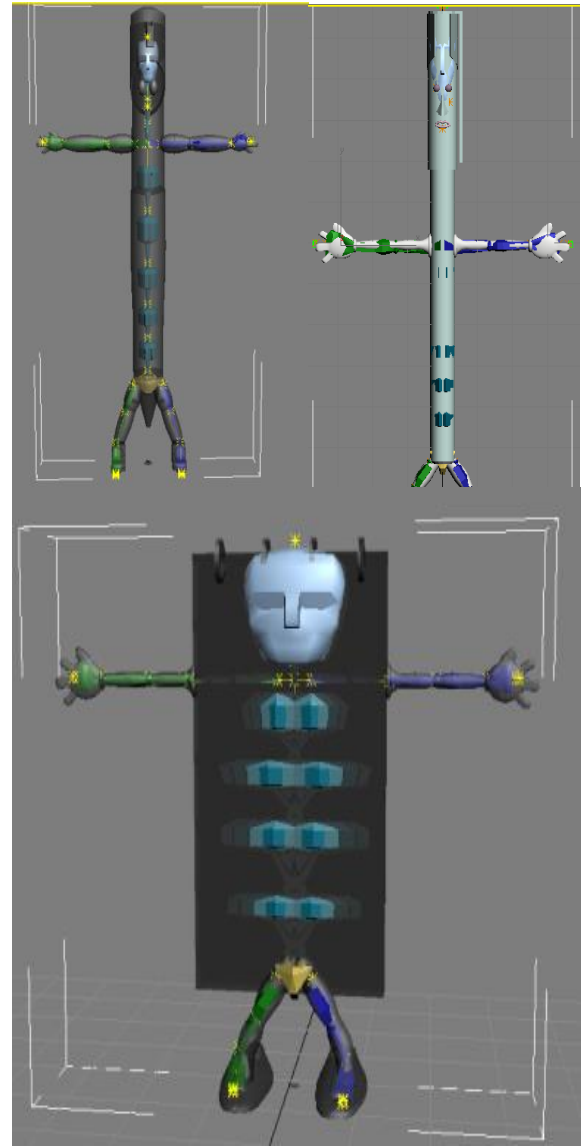

**Figure 19: Rigged and skinned characters**

### 4.4 TEXTURING OF 3D MODELS

In texturing, there are multiple colours and textures used on models of stationary characters to make them look realistic and yet little cartoony and attractive. Figure 20 shows various objects and material assigned to each piece and part of character. Each body part was selected and assigned separate material with special attention given to its colour, specular, reflectivity and ambient attributes.

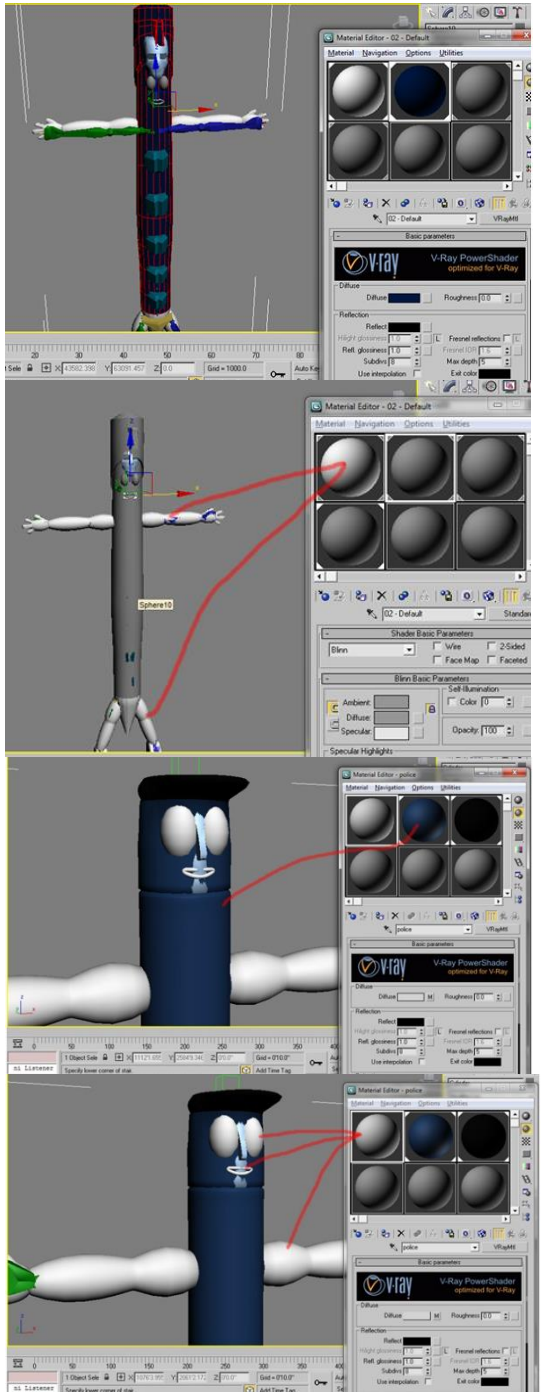

**Figure 20: Material assigned to various pieces of each characters.**

*4.5 ANIMATION*

Animation is a process of bringing virtual objects to life, where artists carefully work with digital objects and characters, manipulating each body part to make them move and appear alive [17]. Technically animation is a sequence of images, created with an impression of movement [18]. Lots of techniques and procedures are used to animate virtual characters [17], [19]. It is, generally, said that a Picture is worth a thousand words, then in this case an animated video is worth ten-thousand word. Tons of information and data can be conveyed by means of moving image called animation, because the eye-brains assembles a sequences of image and interprets them as a continuous movement [20][8].

For animating the characters in this project, most standardised technique of animation called key-frame animation [19] is used, where each body parts of the biped character is selected and moved manually and placed at the desired pose. A key frame was then set at each pose using the pose by pose technique to achieve the desired gesture motion.

*4.6 RENDERING AND COMPOSITION*

For rendering of the final output, 3Ds Max Mental ray rendering system has been used. Mental Ray® is a standalone production quality based 3D rendering technology used by 3Ds max and other 3D Graphics' software to produce high-quality, high-volume and very realistic and life like renderings for complex environments and locations used in film, television, and design visualization projects in less time [21][22]. The final output of the rendering and composition is shown in the next results section.

## V. RESULTS

The final results of the project are very compelling and interesting. The results reflect the intended message with easier way to understand and intuitive manner. The final screenshots of the video are displayed from figures 21 to 32.

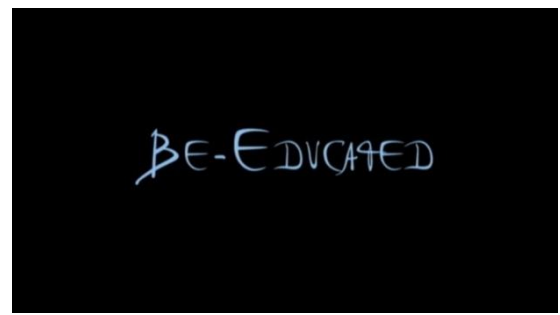

**Figure 21: This is the Starting title of Movie and that title is made in Adobe Photoshop.**

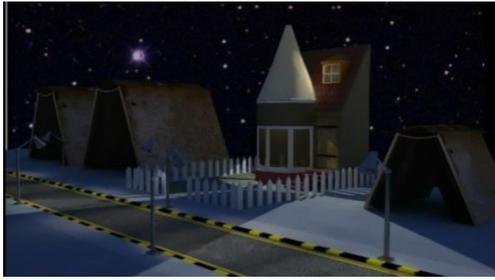

**Figure 22 : This is the starting scene of film, where we are showing the home of ink Pen and the exterior environment in the Night scene.**

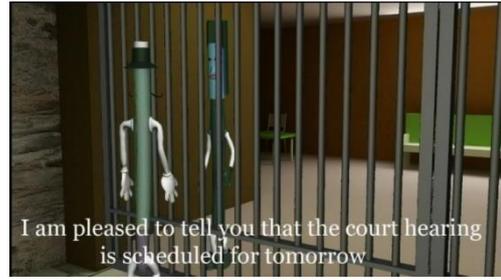

**Figure 26: It's another view of jail, using 2$^{nd}$ camera in 3ds max.**

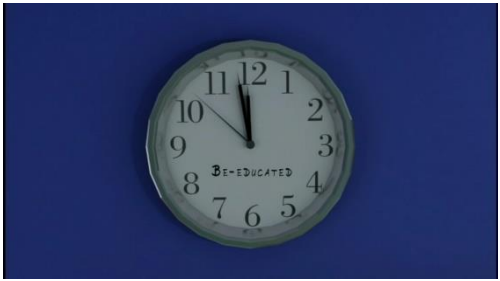

**Figure 23: This is a wall clock which is inside the Room of ink Pen.**

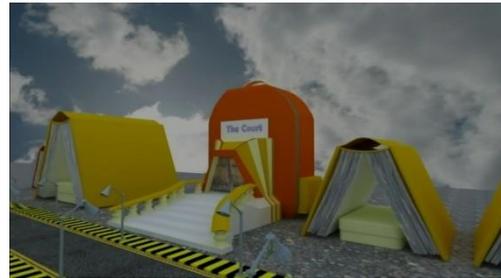

**Figure 27 (right): This is a Court, which is in school bag shape and the exterior environment made up of books, geometry and lamps as a street light.**

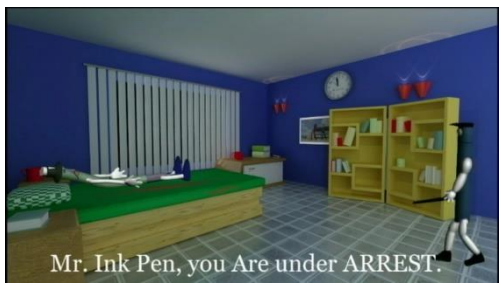

**Figure 24: This is a Room interior of Ink Pen, where he is sleeping on the duster kind of bed, and there are some stationary stuff used as desk, cupboard. These police markers are arresting Mr. Ink pen.**

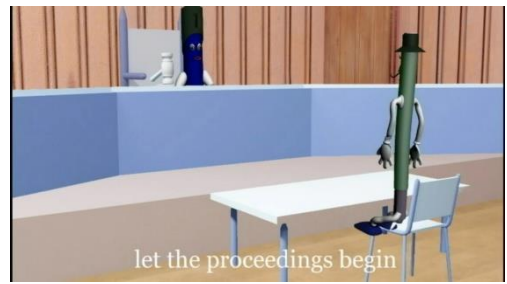

**Figure 28: Court Interior is made up of simple boxes, and pillars of table are in pencil shape.**

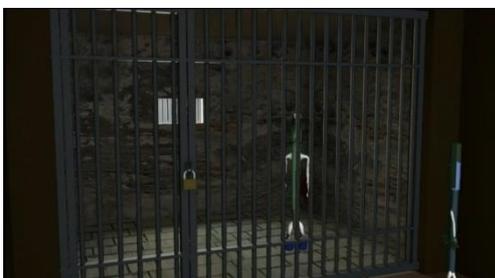

**Figure 25: This is a Jail, where defence Lawyer came to meet Mr. Ink pen. The textures of this environment are designed in Adobe Photoshop and scene is made in 3Ds max.**

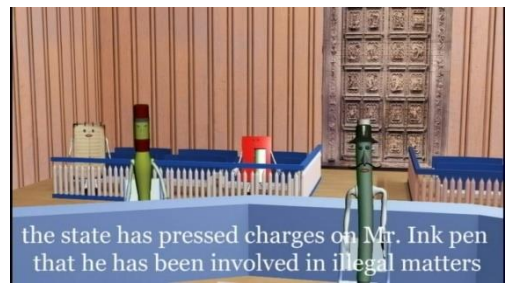

**Figure 29 (right): This is the view from the judge camera, to look up the court interior. Where tables and benches are place made in 3Ds max.**

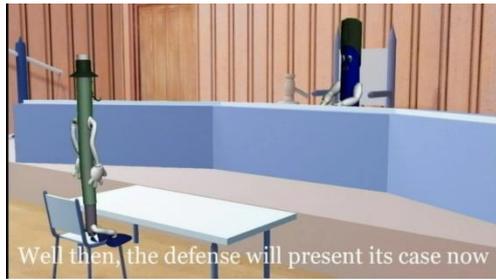

**Figure 30 (left): A view from the female lawyer camera. Showing another side of judge.**

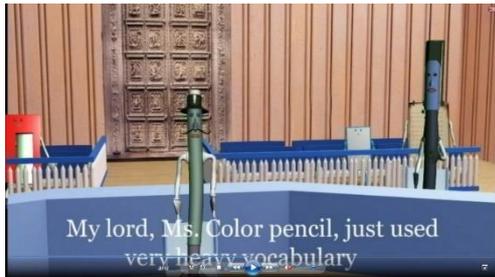

**Figure 31 (right): This is the second camera from judge, viewing the side of defence lawyer.**

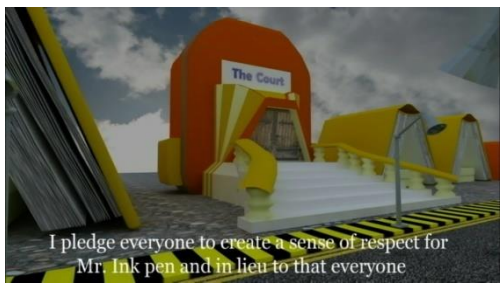

**Figure 32: This is another View of court, another angle. This is the last view.**

## VI CONCLUSION

In this project, a multimedia learning approach was used to educate the students about the importance of education and various tools of education such as pen, marker, pencil, rubber. For this purpose, a 3D animated environment and virtual stationary look-a-like characters were modelled using polygonal modelling techniques in 3D studio Max. These characters were rigged using biped character setup and textured using various materials and shaders. Finally, animation was done based on the story and script initially laid down. The final rendering and later compositing were conducted to produce the broadcast standard video with cognitive learning approach toward the importance of education. Through this 3D animated video, a standard message and lesson has been taught about education and its importance to respect and value the school stationary, without which obtaining education is not possible. The Moral of this Movie is that do not waste your stationary material, use your Pen and Paper for the purpose they are made for. To be a good citizen you have to Be-Educated yourself and for that you need to give value to Pen.

Another advantage of this project is, mostly people in our society are unaware of 3D graphics, in which students are in majority, by implementing this project they will get an idea that what this technology means, what can students make by using 3D graphics as its foundation is all dependable on visualization.

## REFERENCES

[1] Rosenberg, M. L. Statement for the Record on Unintentional Childhood Injury and Death, Assistant Secretary for Legislation (ASL) Department of Health & Human Services (online). Available:http://www,hhs,gov/asl/testify/t980505b.html.

[2] McComas, J. MacKay, M. and Pivik, J. 2002 Effectiveness of Virtual Reality for Teaching Pedestrian Safety. Cyber Psychology &Behavior. 5(3): 185.

[3] Mayer, R. E. (2014). Cognitive theory of multimedia learning. *The Cambridge handbook of multimedia learning*, 43-71.

[4] Mayer, R. E., & Moreno, R. 2002. Animation as an Aid to Multimedia Learning. Educational Psychology Review. 14(1): 87-99.

[5] Mayer, R. E. 1997. Multimedia Learning: Are We Asking the Right Questions? Educational Psychologist. 32(1): 1-19.

[6] Park, B., Plass, J. L., &Brünken, R. (2014). Cognitive and affective processes in multimedia learning. *Learning and Instruction*, *29*, 125-127.

[7] Ann, G. John, N. and Lynne, S. 2005. Using Interactive Multimedia to Teach Pedestrian Safety: An Exploratory Study. Am J Health Behav. 29(5): 435-442.

[8] Rawi, N. A., Mamat, A. R., Deris, M. S. M., Amin, M. M., & Rahim, N. (2015). A Novel Multimedia Interactive Application to Support Road Safety Education Among Primary School Children In Malaysia. *JurnalTeknologi*, *77*(19).

[9] Ahmed, I., &Janghel, S. (2015). 3D Animation: Don't Drink and Drive. International Journal of u-and e-Service, Science and Technology, 8(1), 415-426.

[10] Jago, M. (2014). Adobe Premiere Pro CC Classroom in a Book (2014 release). Adobe Press.

[11] Team, A. C. (2013). *Adobe Audition CC Classroom in a Book*. Adobe Press.

[12] Kaboli, A., Jalili, T., Sorkhdan, N. A. G., Fazeli, Z., &Arghande, N. (2016). The role of 3dmax in architecture. *Journal of Current Research in Science*, (1), 271.

[13] Chandler, M., Podwojewski, P., Amin, J., & Herrera, F. (2014). 3ds Max Projects: A Detailed Guide to Modeling, Texturing, Rigging, Animation and Lighting.

[14] Bhatti, Z., Shah, A., Waqas, A., Karbasi, M., Mahesar, A. W. (2015) A Wire Parameter and Reaction Manager based Biped Character Setup and Rigging Technique In 3Ds Max for Animation. International Journal of Computer Graphics & Animation (IJCGA) Vol.5, No.2. April 2015

[15] Z. Bhati, A. Shah, A. Waqas, H. Abid, and M. Malik, "Template based Procedural Rigging of Quadrupeds with Custom Manipulators," in International Conference on Advanced Computer Science Applications and Technologies, 2013, pp. 259–264

[16] Z. Bhatti and A. Shah, "Widget based automated rigging of bipedal character with custom manipulators," Proc. 11th ACM SIGGRAPH Int. Conf. Virtual-Reality Contin. its Appl. Ind. -VRCAI '12, p. 337, 2012

[17] Bhatti, Z., Shah, Asadullah, K. M., and Mahesar, W. 2013. Expression driven trigonometric based procedural animation of quadrupeds. In Proceedings of the International Conference on Informatics and Creative Multimedia 2103 (ICICM'13), pp.104,109, 4-6 Sept. 2013 doi:10.1109/ICICM.2013.25. IEEEXplore. UTM, Kuala Lumpur

[18] Savage, T. M. and Vogel, K. E. 2014. An Introduction to Digital Multimedia. 2nd Edition: Jones & Bartlett Learning Publication.

[19] Bhatti, Z., Shah, A., &Shahidi, F. (2013, November). Procedural model of horse simulation. In Proceedings of the 12th ACM SIGGRAPH International Conference on Virtual-Reality Continuum and Its Applications in Industry (VRCAI '13). ACM, New York, NY, USA, 139-146. DOI=10.1145/2534329.2534364 http://doi.acm.org/10.1145/2534329.2534364 ACM.

[20] MohdIzani, Aishah', Ahmad RaifiEshaq' and Norzaiha. 2003. Keyframe Animation and Motion Capture for Creating Animation: A Survey and Perception from Industry People. Student Conference on Research and Development (SCOReD) 2003 Proceedings, Putrajaya, Malaysia.

[21] O'Connor, R. (2015). *Beginner's Guide To Mental Ray and Autodesk Materials In 3ds Max 2016*. CreateSpace Independent Publishing Platform.

[22] Driemeyer, T. (2013). *Rendering with mental ray®* (Vol. 1). Springer.